\documentclass[twocolumn,twoside,5p,preprint]{elsarticle}
\biboptions{sort&compress}

\usepackage[utf8]{inputenc}
\usepackage[english]{babel}
\usepackage{amssymb,amsthm,amsmath,amstext,amsbsy,amsopn}
\usepackage{bbm}
\usepackage{graphicx}
\usepackage{isotope}
\usepackage{float}
\usepackage[dvipsnames]{xcolor}

\usepackage{hyperref}
\hypersetup{
    colorlinks = true,
    citecolor  = blue,
    linkcolor  = blue,
    urlcolor  = black
}

\definecolor{plot1}{rgb}{0.86, 0.08, 0.24}
\definecolor{plot2}{rgb}{0.25, 0.41, 0.88}
\definecolor{plot3}{rgb}{1.0, 0.55, 0}
\definecolor{plot4}{RGB}{61,153,86}

\newcommand{\la}{\langle}
\newcommand{\ra}{\rangle}

\newcommand{\lang}{\ensuremath{l}}
\newcommand{\langpr}{\ensuremath{l'}}

\newcommand{\ai}{{\emph{ab initio}}}

\newcommand{\RSVD}{\ensuremath{R_\text{SVD}}}

\newcommand{\MeV}{\ensuremath{\text{MeV}}}
\newcommand{\fm}{\ensuremath{\text{fm}}}
\newcommand{\fmi}{\ensuremath{\text{fm}^{-1}}}

\newcommand*{\nxlo}[1]{N${}^{#1}$LO}

\newcommand{\compression}[2]{\ensuremath{C^{#1}_{#2}}}

\begin{document}

\allowdisplaybreaks

\title{Low-rank matrix decompositions for \ai{} nuclear structure}

\author[tud,emmi,mpik]{A.~Tichai}
\ead{alexander.tichai@physik.tu-darmstadt.de} 
\author[tud,emmi]{P.~Arthuis}
\ead{parthuis@theorie.ikp.physik.tu-darmstadt.de}
\author[tud,emmi]{K.~Hebeler}
\ead{kai.hebeler@physik.tu-darmstadt.de}
\author[tud,emmi]{M.~Heinz}
\ead{mheinz@theorie.ikp.physik.tu-darmstadt.de}
\author[tud,emmi]{J.~Hoppe}
\ead{jhoppe@theorie.ikp.physik.tu-darmstadt.de}
\author[tud,emmi,mpik]{A.~Schwenk}
\ead{schwenk@physik.tu-darmstadt.de}

\address[tud]{Technische Universit\"at Darmstadt, Department of Physics, 64289 Darmstadt, Germany}
\address[emmi]{ExtreMe Matter Institute EMMI, GSI Helmholtzzentrum f\"ur Schwerionenforschung GmbH, 64291 Darmstadt, Germany}
\address[mpik]{Max-Planck-Institut f\"ur Kernphysik, Saupfercheckweg 1, 69117 Heidelberg, Germany}

\begin{abstract}
The extension of \ai{} quantum many-body theory to higher accuracy and larger systems is intrinsically limited by the handling of large data objects in form of wave-function expansions and/or many-body operators. 
In this work we present matrix factorization techniques as a systematically improvable and robust tool to significantly reduce the computational cost in many-body applications at the price of introducing a moderate decomposition error.
We demonstrate the power of this approach for the nuclear two-body systems, for many-body perturbation theory calculations of symmetric nuclear matter, and for non-perturbative in-medium similarity renormalization group simulations of finite nuclei. Establishing low-rank expansions of chiral nuclear interactions offers possibilities to reformulate many-body methods in ways that take advantage of tensor factorization strategies.
\end{abstract}


\maketitle

\section{Introduction}

The computational requirements of quantum many-body theory grow rapidly due to the need to handle large data objects, either in the form of wave-function expansions or operator matrix elements encoding the dynamics of the interacting particles.
In particular, extending the scope of \ai{} nuclear theory to heavier systems comes at the cost of enormous storage requirements when incorporating three-body matrix elements that are mandatory for a quantitative reproduction of nuclear phenomena (see, e.g., Refs.~\cite{Hebe15ARNPS,Hebe203NF}).
Over the last years state-of-the-art applications provided converged many-body calculations up to mass number $A\approx 100$ based on chiral two- and three-nucleon interactions~\cite{Morr17Tin,Herg20review}.
While first exploratory studies targeted selected heavier nuclei~\cite{Arthuis2020a,Miyagi2021}, a systematic and accurate description of nuclear observables in heavy systems is still beyond the scope of even the most advanced \ai{} calculations.

There are also ongoing efforts in the nuclear theory community to include deformation effects in many-body calculations from first principles~\cite{Duguet2014su2,Yao18IMGCM,Novario2020a,Frosini2021}.
The use of symmetry-broken many-body expansions increases the memory requirements tremendously even at moderate truncation levels in the many-body basis expansion.
Therefore, future applications will eventually require novel formats for the representation of many-body wave functions and/or operators when working in non-spherical computational bases.

\begin{figure*}[t!]
    \centering
    \includegraphics[width=\textwidth]{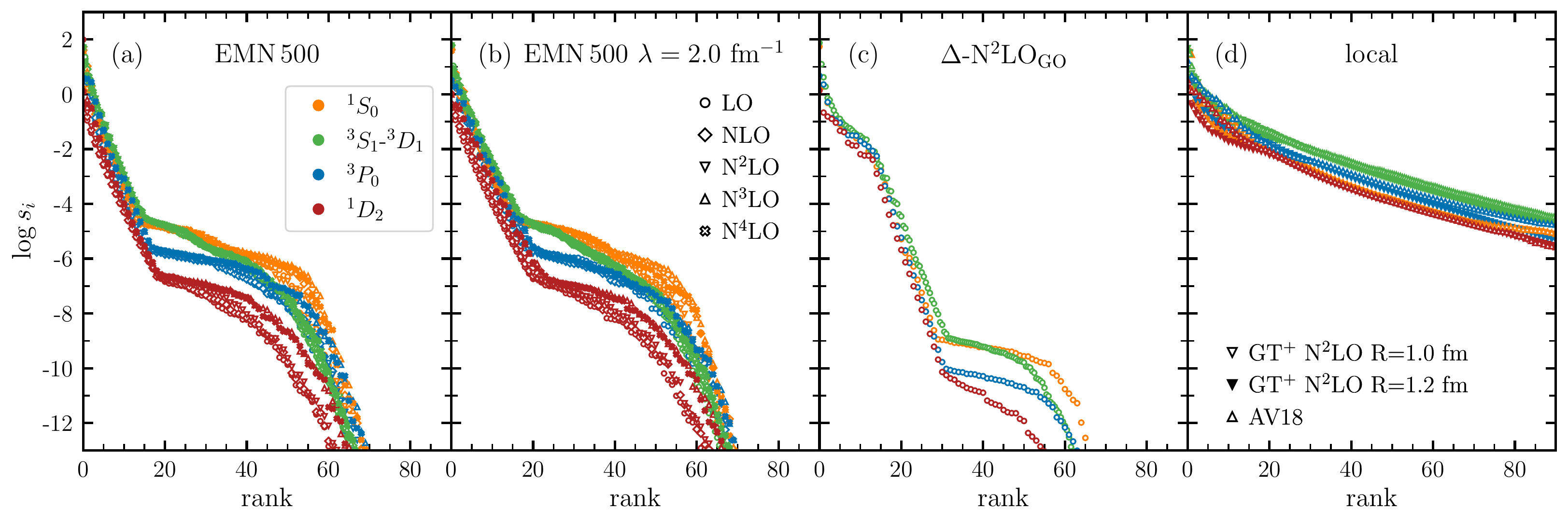}
    \caption{Singular values of the EMN~500~\cite{Ente17EMn4lo} (a-b), $\Delta$-\nxlo{2}$_\text{GO}$~\cite{Jian20N2LOGO} (c), and local N$^2$LO (GT+)~\cite{Geze13QMCchi} and AV18~\cite{Wiri95AV18} (d) two-nucleon potentials as a function of SVD rank. Different colors correspond to different partial-wave channels.
    In panels (a) and (b) we consider potentials at multiple chiral orders from LO to N$^4$LO indicated by the different shapes.
    The potentials used in panel (b) differ from those in panel (a) by an SRG evolution to a lower resolution scale $\lambda = 2.0 \, \fmi$.
    In the coupled deuteron channel the rank was divided by two in order to account for the effective doubling of the matrix dimension.
    }
    \label{fig:svals}
\end{figure*}

Efforts to lower the computational expense are often focused on a restriction of the configuration space used in the many-body method, e.g., using importance sampling in configuration interaction calculations~\cite{Roth09ImTr} or in self-consistent Green's function theory~\cite{Porro2021}, where the importance of many-body configurations for a given observable is gauged and states deemed unimportant are discarded from the calculations.
While the memory requirements for storing wave functions or correlation expansions can be significantly lowered using this strategy, the underlying fundamental limitations for storing many-body operators still persist.
One possible path to lowering the memory footprint is the use of tensor factorization (TF) techniques for high-dimensional data objects by approximating them as sums/products of low-dimensional ones~\cite{Kolda2009}.
In proof-of-principle applications for nuclei, it was shown that TF approaches yield significant compression ratios when applied in single-particle harmonic oscillator (HO) and Hartree-Fock (HF) bases~\cite{Tich19THC} as well as in quasi-particle extensions~\cite{Tichai2019pre}. These applications employed compression techniques on the final single-particle matrix elements obtained at the end of a series of basis transformations.
Ideally, one would however apply the compression to the initial momentum space representation
and reformulate the transformations to the single-particle basis in terms of decomposition factors to fully take advantage of the compression.

In the past, singular value decomposition techniques have also been applied in sensitivity studies of energy density functionals~\cite{Bertsch2005}, for shell-model interactions~\cite{Brow06USD,Jaganathen2017,Fox2020}, in electrodisintegration studies~\cite{More17sdde}, or as tools for generalized eigenvalue problems~\cite{Robledo2011oct}.
Moreover, low-rank interactions have also been obtained based on expansions in Weinberg eigenvalues~\cite{Bogn06bseries,Rama07WEVPair,Hopp17WeinEVAn}.

In this Letter, TFs are applied directly to the momentum-space representation of nuclear interactions obtained from chiral effective field theory (EFT)~\cite{Epel09RMP,Mach11PR,Geze13QMCchi,Ente17EMn4lo,Hebe203NF,Jian20N2LOGO}.
The decomposed matrix representations are then employed in different nuclear calculations to benchmark the performance in two-body applications, symmetric nuclear-matter calculations, and for ground-state properties of closed-shell nuclei.
Having systematic control over the decomposition error complements the \ai{} character established by employing nuclear interactions based on the symmetries of the underlying theory of quantum chromodynamics (QCD) combined with systematically improvable many-body methods.

\section{Matrix decompositions}
\label{sec:matdec}

The key computational tool employed in this work is the singular-value decomposition (SVD) of a matrix $V$,
\begin{align}
V = L \Sigma R^T \,,
\end{align}
where $L$ ($R$) denotes the left (right) singular vectors, and the diagonal matrix $\Sigma = \text{diag}(s_i)$ contains the set of non-negative singular values $s_i$ in order of decreasing magnitude.
The columns of the left (right) singular vectors $L$ ($R$) are orthonormal, i.e., $L$ ($R$) is a unitary matrix.
Matrix factorizations provide a basis for designing low-rank approximations by identifying and discarding unimportant information from small singular values in the decomposition. 
In the following, the truncated SVD is employed
\begin{align}
    \tilde V \equiv \tilde L \tilde \Sigma \tilde R^T \,,
    \label{eq:tsvd}
\end{align}
where all singular values beyond the decomposition rank $\RSVD$ are discarded, i.e.,
\begin{align}
    \tilde{\Sigma} \equiv \text{diag}(s_1, \ldots, s_{\RSVD}, 0, \ldots, 0) \, 
    ,
\end{align}
such that only the first $\RSVD$ columns of $L$ (rows of $R$) need to be stored.
According to the Eckart-Young theorem the truncated SVD $\tilde V$ provides the best rank-$\RSVD$ approximation to $V$ in the sense of minimizing
\begin{align}
    \Vert \Delta V \Vert_\text{F}^2 \equiv \Vert V - \tilde V \Vert_\text{F}^2  \,,
    \label{eq:froerror}
\end{align}
where $\Vert \cdot \Vert_\text{F}$ denotes the Frobenius norm.
The SVD is a particularly versatile factorization ansatz because it naturally extends to non-Hermitian and non-square matrices, a feature not shared by some other common matrix decompositions such as the eigenvalue or Cholesky decompositions.
For two-body interactions, truncating the SVD at rank $\RSVD=1$ produces a separable approximation to the potential, but for $\RSVD > 1$ the truncated SVD differs from higher-rank separable approximations.

Comparing the storage requirements of the full interaction matrix ($N^2$ with matrix dimension $N$) to the storage requirements of the truncated factors ($2N\RSVD + \RSVD$) on the right-hand side of Eq.~\eqref{eq:tsvd} allows one to define the compression
\begin{align}
 \compression{N}{\RSVD} \equiv \frac{N^2}{2 N \RSVD + \RSVD} \, 
\end{align}
as the ratio of the initial storage needed divided by the storage needed after factorization.
The higher the compression the higher the computational savings.
However, higher compression typically comes at the price of inducing higher decomposition errors such that in practice a balance has to be found.

\section{Low-rank interactions}

Before targeting nuclear observables we start by investigating low-rank decompositions of two-nucleon ($NN$) interaction matrix elements in a momentum-space partial-wave basis
\begin{equation}
\la p \, (\lang S) J T M_T | V_{NN} | p' \, (\langpr S) J T M_T \ra\,,
\end{equation}
with the final and initial orbital angular momenta  $\lang$ and $\langpr$, the two-body spin $S$, the two-body total angular momentum $J$, the two-body isospin $T$ with projection $M_T$, and the absolute values of the outgoing and incoming relative momenta $p$ and $p'$. In each partial-wave channel, the $NN$ potential is represented using $N=100$ momentum mesh points up to $p_{\text{max}}=p_{\text{max}}'=6.0\,\fm^{-1}$.
In this work, we focus on the family of chiral two-nucleon potentials given up to next-to-next-to-next-to-next-to-leading order (\nxlo{4}) from Entem, Machleidt, and Nosyk (EMN)~\cite{Ente17EMn4lo} for the cutoff 500\,MeV (EMN~500) and consider various other potentials for comparison~\cite{Wiri95AV18,Geze13QMCchi,Jian20N2LOGO}.
In some cases we use the free-space similarity renormalization group (SRG)~\cite{Bogn07SRG} to soften the potentials.

Figure~\ref{fig:svals} displays the singular values for selected partial-wave channels $^{2S+1}\lang_{J}$.
Panel (a) shows the singular values for the EMN~500 potentials,
and panel (b) uses the same potential SRG-evolved to a resolution scale of $\lambda=2.0\,\fm^{-1}$.
In both cases, we show results for different chiral orders ranging from leading order (LO) to \nxlo{4}.
Panel (c) is for the $\Delta$-\nxlo{2}$_\text{GO}$ interaction~\cite{Jian20N2LOGO}.
Panel (d) uses local N$^2$LO (GT+) potentials~\cite{Geze13QMCchi} (for cutoffs $R=1.0$ and $1.2\,\fm$) and the Argonne AV18 potential~\cite{Wiri95AV18}.
Different colors correspond to selected $S, P, D$ uncoupled partial waves as well as the coupled deuteron channel.

For the EMN~500 interaction we observe a rapid fall-off of singular values down to magnitudes below $10^{-4} - 10^{-5}$ from $\RSVD \approx 20$ on.
The behavior is the same in all partial-wave channels while the rank is effectively doubled in the deuteron channel due to the coupling of $S$-$D$ blocks. Moreover, partial waves with higher angular momenta generally have smaller singular values.
All of the above features are independent of the chiral order, and no systematic differences arise at different chiral orders.
This shows that the low-rank properties of chiral interactions are not spoiled by the presence of more complicated operator structures that enter with higher-order pion exchanges or higher-order short-range interactions.
Comparing panel (a) and (b) indicates that the SRG-evolution does not affect the singular value behavior as well.
Although momentum-space matrix elements are different, the SRG has little effect on the SVD fall-off.

Panel (c) shows results for the $\Delta$-\nxlo{2}$_\text{GO}$ interaction that explicitly incorporates $\Delta$ isobar degrees of freedom.
Interestingly, the decrease of the singular values is initially milder than for the EMN potentials, while it is followed by a very fast decrease, down to a plateau around magnitudes of $10^{-9} - 10^{-10}$ for $\RSVD \gtrsim 30$. This fall-off is more pronounced than for the EMN~500 potentials. We suspect that these differences in the characteristics of the singular values can be traced back to the different fitting protocols for these two families of potentials. While the $\Delta$-\nxlo{2}$_\text{GO}$ potential was designed for applications to medium-mass nuclei and the fit to phase shifts focused on laboratory energies up to about 125\,MeV, the EMN interactions can accurately reproduce phase shift data up to 300\,MeV~\cite{Ente17EMn4lo}.

Panel (d) displays three different local $NN$ interactions, two chiral ones at \nxlo{2} with cutoffs $R=1.0$ and $1.2\,\fm$ and the AV18 potential. We observe that for these local potentials, the decrease of singular values is slower than for the non-local cases (a) to (c).
Even for rank $\RSVD > 50$ the singular values are of significant size compared to those of the non-local EMN~500 interaction.
We note that the present SVD in momentum space is probably not optimal for local interactions, as the SVD in momentum space entails non-local features.
It will be interesting to explore SVD strategies in coordinate space, but this is left to future work.

\begin{figure}[t!]
    \centering
    \includegraphics[width=\columnwidth]{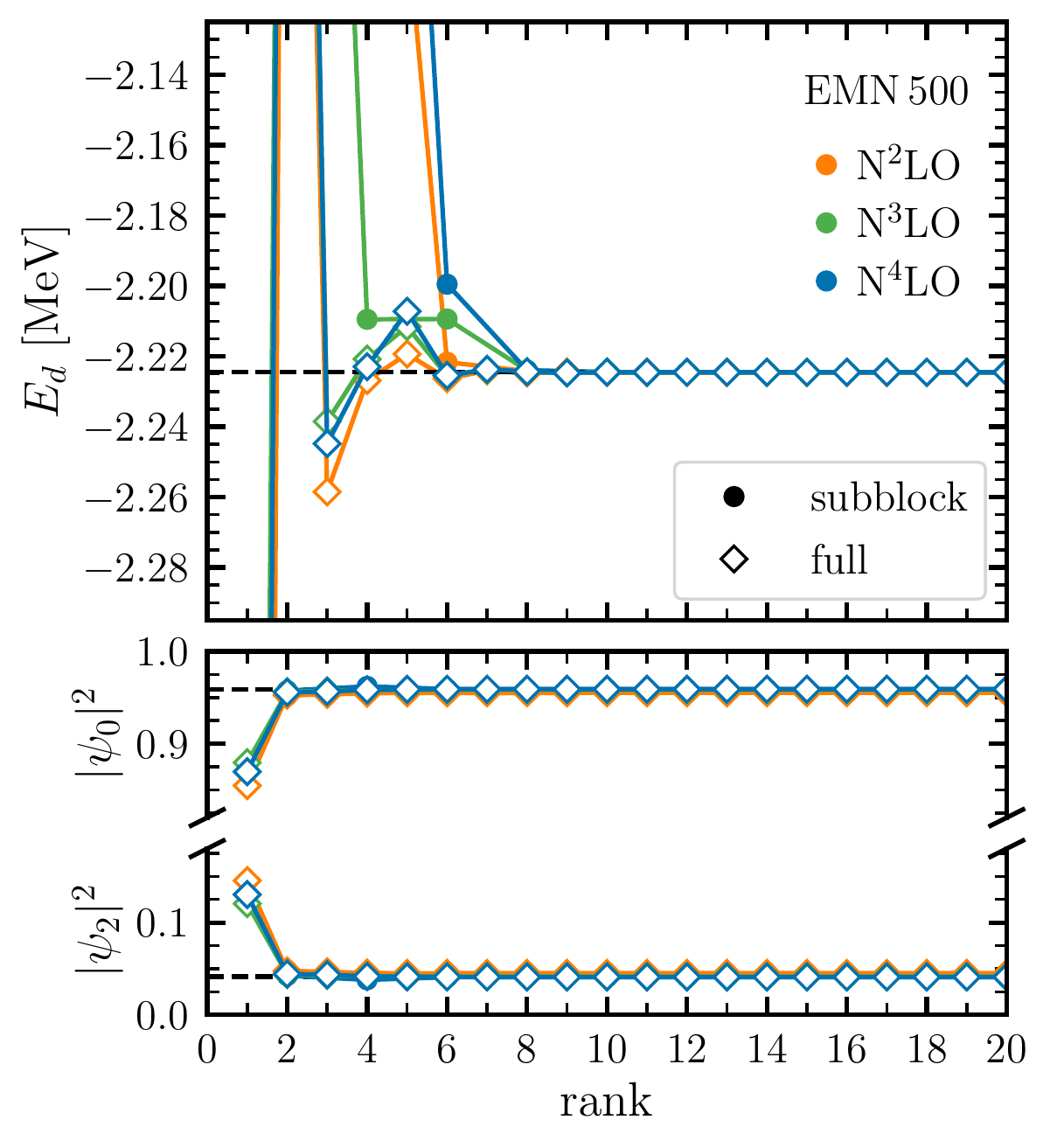}
    \caption{Top panel: Ground-state energy of the deuteron as a function of SVD rank. The SVD is either performed in the ``full'' (diamonds) or the ``subblock'' (circles) matrix of the coupled channel (see text for details). Bottom panel: $S$- and $D$-state normalizations as a function of SVD rank when following the ``full'' matrix approach. All results are for the EMN~500 $NN$ interaction, while the color scheme indicates the chiral orders from \nxlo{2} to \nxlo{4}.}
    \label{fig:deuteron}
\end{figure}

\section{Two-nucleon system}

\begin{figure*}[t!]
    \centering
    \includegraphics[width=\textwidth]{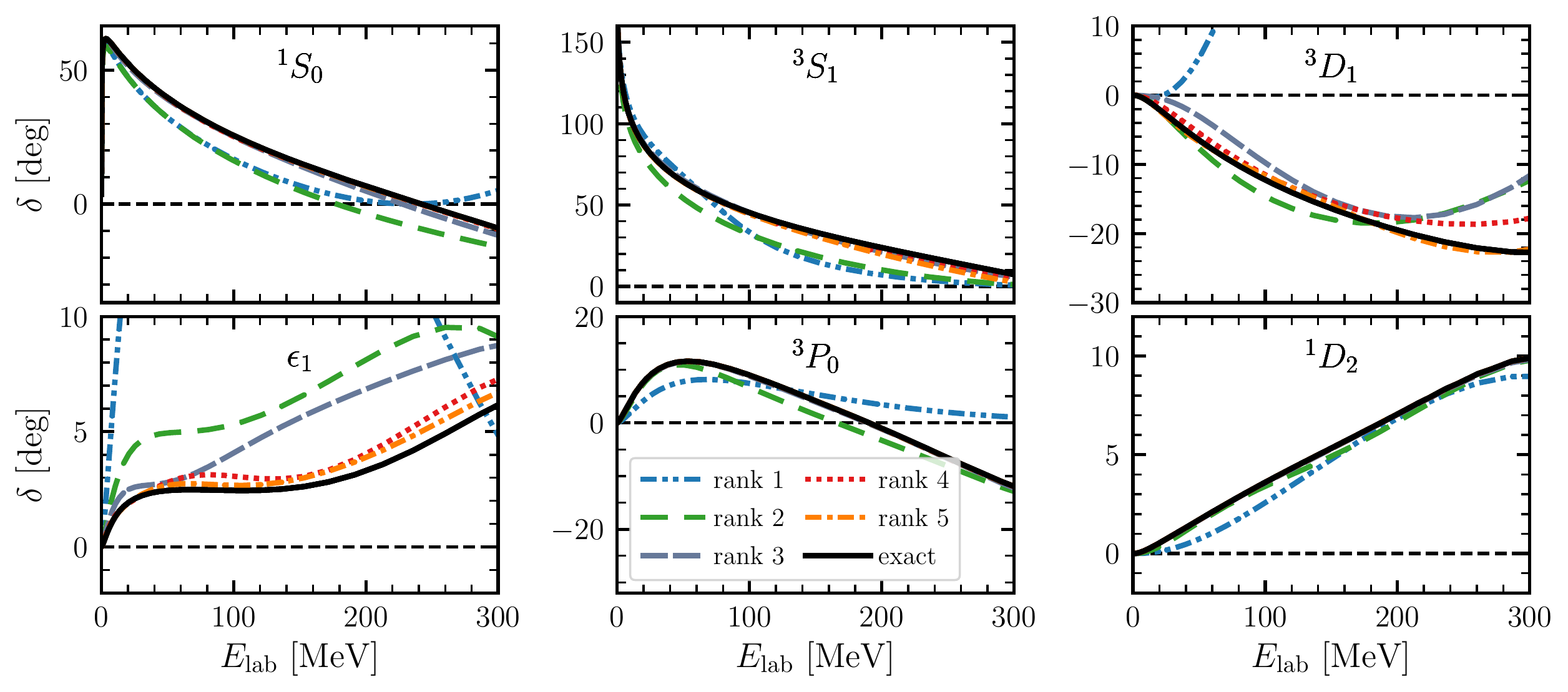}
    \caption{Phase shifts of the EMN~500 $NN$ interaction at \nxlo{3} for selected partial-wave channels.
    The bottom left panel shows the mixing angle $\epsilon_1$ for the coupled $^3S_1$-$^3D_1$ channel.
    Exact results are shown in black while those from low-rank approximations are indicated by the colored lines.}
    \label{fig:phaseshifts}
\end{figure*}

As the simplest observable, we target the deuteron ground-state energy obtained by diagonalizing the Hamiltonian in the coupled $^3S_1$-$^3D_1$ channel.
In a coupled partial-wave channel the SVD can be either performed on the $2N\times 2N$ block matrix built from the four partial channels (``full'') or performed individually on each of the four $N\times N$ submatrices (``subblock'').
The feasibility of the second option features the flexibility of the SVD, as it may be performed for non-Hermitian matrices like the off-diagonal submatrices in spin-triplet channels.
Figure~\ref{fig:deuteron} shows the deuteron ground-state energy as a function of SVD rank for the ``full'' and ``subblock'' matrix approach using the EMN~500 interaction at \nxlo{2}, \nxlo{3}, and \nxlo{4}. For both approaches the exact binding energy is reproduced at the sub-percent level at $\RSVD \approx 6$, and beyond $\RSVD=10$ differences are well below the keV level. Note that for the EMN interactions, the exact binding energy is identical at all chiral orders since the low-energy couplings of the potentials are fit to reproduce the deuteron binding energy.

In addition, the lower panel of Fig.~\ref{fig:deuteron} shows the $S$- and $D$-state normalizations of the deuteron wave function as a function of SVD rank using the ``full'' approach.
The quantitative reproduction of the small $D$-wave admixture due to the presence of non-central tensor forces is well captured already at very low $\RSVD$ such that the ground-state energy and the two-body wave function can be accurately approximated from low-rank potentials at $\RSVD \sim 5$.

To further benchmark low-rank nuclear potentials in the two-body system, we investigate two-nucleon scattering phase shifts obtained by solving for the $T$ matrix satisfying the Lippmann-Schwinger integral equation for the truncated SVD interaction $\tilde V$
\begin{align}
    &\la p \alpha | T| p'  \alpha' \ra =
    \la p \alpha | \tilde V | p'  \alpha' \ra 
    \notag \\ 
    & \quad +
    \frac{2}{\pi} \sum_{\alpha''} \int_0^\infty dq \, q^2 \,
    \frac{ \la p \alpha| \tilde V | q \alpha'' \ra \la q \alpha'' | T | p' \alpha' \ra }{E - \frac{q^2}{m} + i \epsilon} \, ,
    \label{eq:LS}
\end{align}
where the $T$ matrix is evaluated right-on-shell, i.e., $E=p'^2/2m$, and the
combined indices $\alpha, \alpha', \alpha''$ include all partial-wave quantum numbers, but only differ in $l, l', l''$.

In Fig.~\ref{fig:phaseshifts} we compare $NN$ scattering phase shifts for the EMN~500 interaction at \nxlo{3} up to laboratory energies of $300\,\MeV$ using the same set of partial-wave channels as in Fig.~\ref{fig:svals}.
For very small SVD ranks, i.e., $\RSVD = 1$ or $2$, scattering phase shifts obtained from the low-rank potentials show significant deviations compared to the exact calculations for the shown energy range.
When more components are included in the decomposition the deviation to the exact results is systematically reduced, and at ranks $\RSVD \gtrsim 5$ there is a very quantitative reproduction of the exact phase shifts over the entire energy regime.
The difference to the exact results for $\RSVD=5$ is below $\Delta \delta < 1^\circ$ in all partial-wave channels, with only noticable differences at higher $E_\text{lab}$.
In the deuteron channel we also show the mixing angle $\epsilon_1$, which is sensitive to all contributing subblocks in the $^3S_1$-$^3D_1$ channel.
For $E_\text{lab} > 150\,\MeV$ the difference to the exact mixing angle is about $\Delta \epsilon_1 =  1^\circ$ for $\RSVD=5$. When increasing the rank to $\RSVD=15$ (not shown) virtually exact results are obtained.

\section{Nuclear matter}

Next we test low-rank nuclear interactions as input to nuclear-matter calculations in the density regime accessible by chiral effective field theory~\cite{Hebe11fits,Dris17MCshort,Leonhardt2020}.
The energy per volume $E/V$ of symmetric nuclear matter is evaluated up to second-order in many-body perturbation theory (MBPT) using only $NN$ interactions,
\begin{align}
    \frac{E^{(1)}}{V} &= \frac{1}{2}
    \prod_{i=1}^2 \biggl( \text{Tr}_{\sigma_i,\tau_i} 
    \int \frac{d \mathbf{k}_i}{(2\pi)^3} \biggr) \nonumber \\
    &\quad \quad \times \la 12| \tilde V_{NN} (1-P_{12})| 12 \ra n_1 n_2 
    \,, \\[2mm]
    \frac{E^{(2)}}{V} &= \frac{1}{4}
    \prod_{i=1}^4 \biggl( \text{Tr}_{\sigma_i,\tau_i} 
    \int \frac{d \mathbf{k}_i}{(2\pi)^3} \biggr) \frac{|\la 12| \tilde V_{NN} (1-P_{12}) | 34 \ra|^2}{\varepsilon_1 + \varepsilon_2 - \varepsilon_3 - \varepsilon_4} \nonumber \\
    &\quad \quad \times n_1 n_2 \bar n_3 \bar n_4 \, (2\pi)^3 \delta(\mathbf{k}_1 + \mathbf{k}_2 - \mathbf{k}_3 - \mathbf{k}_4) \,,
\end{align}
with the short-hand notation $|1\ra\equiv |\mathbf{k}_1 \sigma_1 \tau_1 \ra$ that gathers all single-particle state labels. The trace is over spin and isospin, the integrations are performed over single-particle momentum states, $n_i$ are zero-temperature Fermi-Dirac functions, and $\bar n_i \equiv 1 - n_i$.
The energy contributions at first and second order are calculated in 
a partial-wave-decomposed form (see, e.g., Ref.~\cite{Hebe10nmatt}) with the EMN~500 $NN$ interaction.
The results presented here are consistent with the ones obtained using the Monte-Carlo framework from Ref.~\cite{Dris17MCshort}.

\begin{figure}[t!]
    \centering
    \includegraphics[width=\columnwidth]{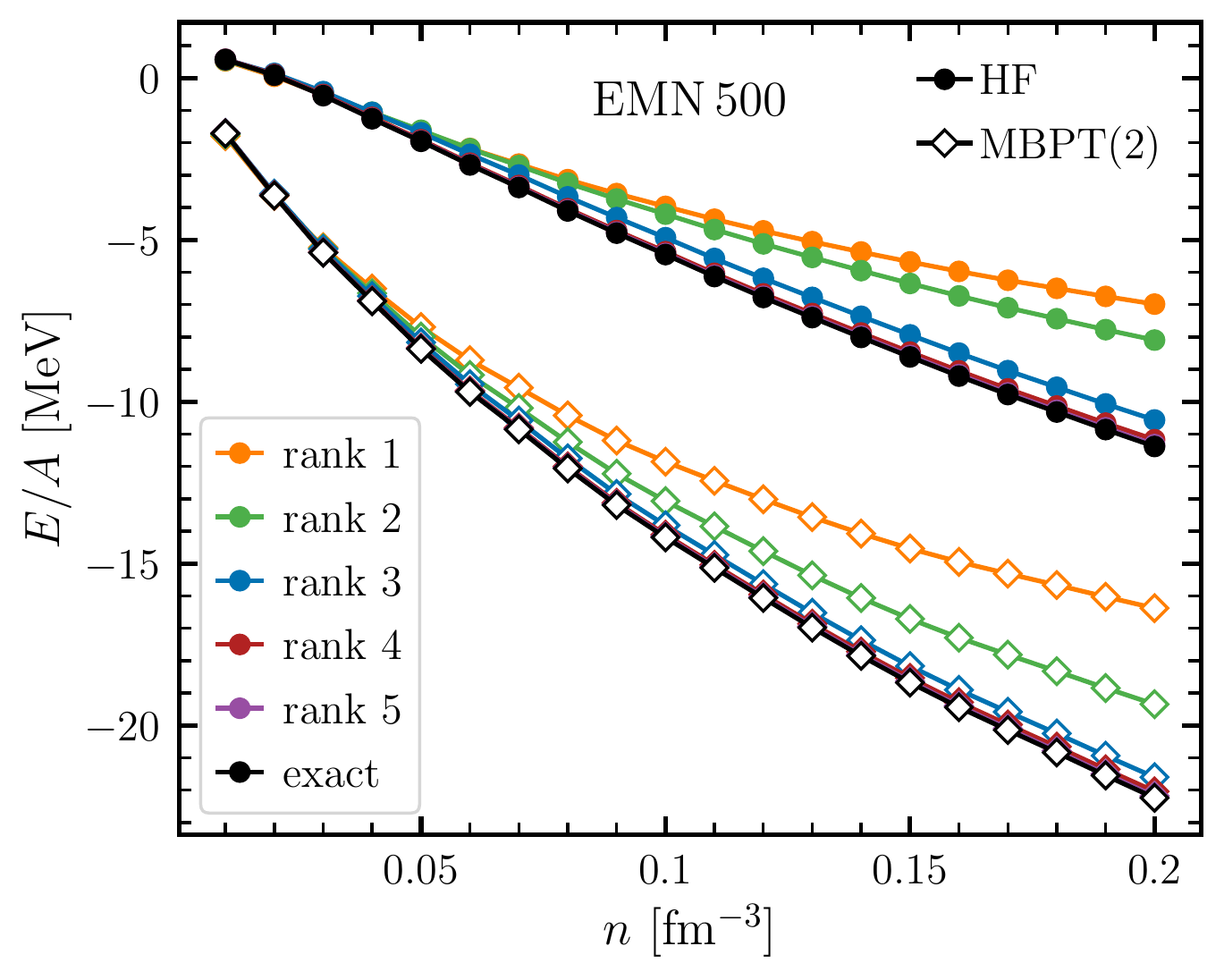}
    \caption{Energy per particle $E/A$ as a function of density $n$ of symmetric nuclear matter at the Hartree-Fock (HF) level (full circles) and at second-order MBPT (open diamonds). Different colors correspond to different SVD ranks in the decomposition of the EMN~500 $NN$ interaction.}
    \label{fig:SNM}
\end{figure}

For the decomposition we follow the ``subblock'' matrix approach and decompose each partial-wave channel separately using the same rank in the SVD format.
Note that in a more refined approach $\RSVD$ can be dynamically chosen by setting a precision threshold $s_\text{min}$ and discarding numerical values $s_i < s_\text{min}$.
This may efficiently adapt to stronger suppression of singular values in partial-wave channels with high angular momenta that are less important for the MBPT contributions.
We will investigate this strategy more in the future.

\begin{figure}[t!]
    \centering
    \includegraphics[width=\columnwidth]{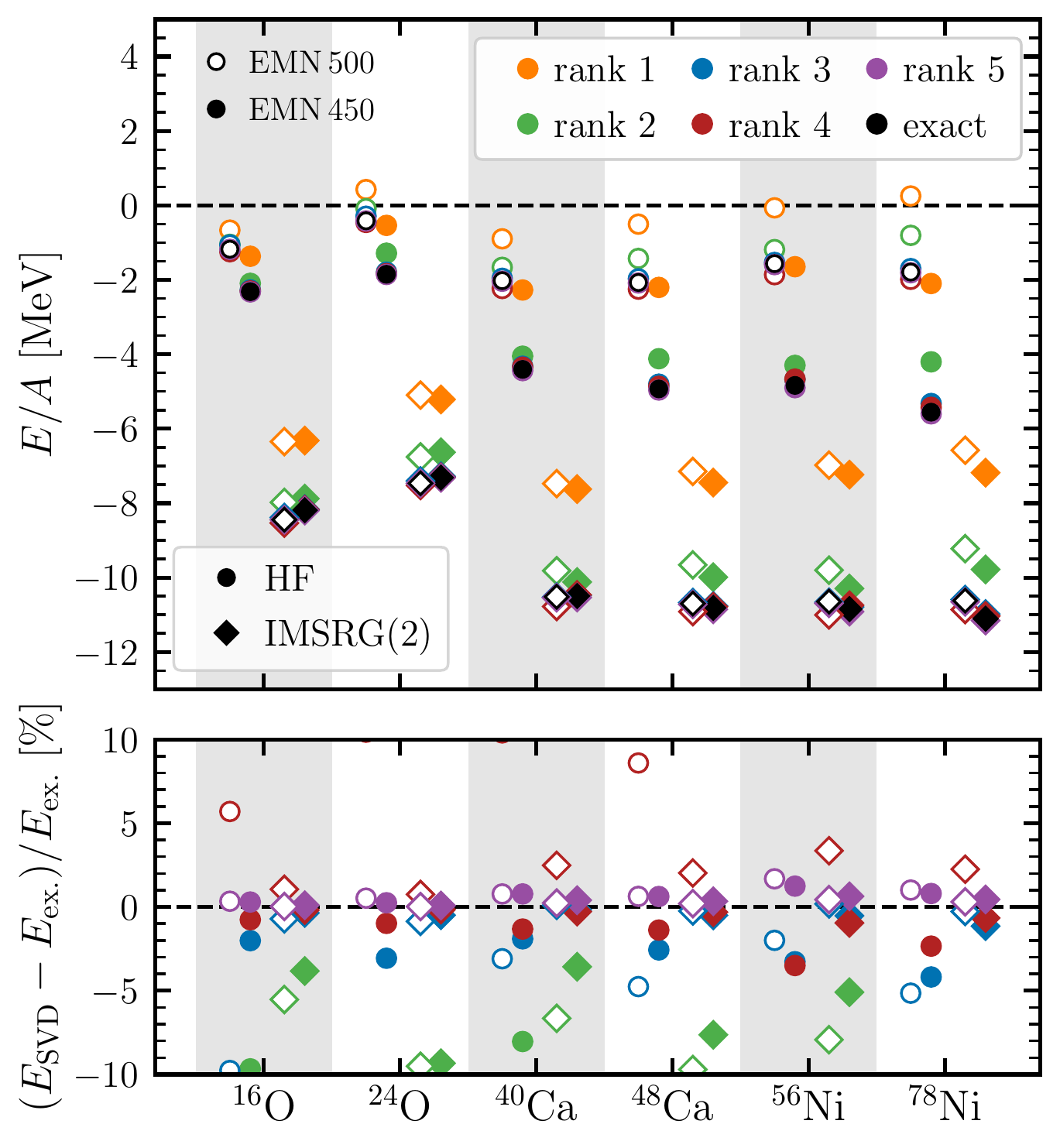}
    \caption{Top panel: Ground-state energy per particle of selected closed-shell nuclei at the HF level (circles) and using the IMSRG(2) approximation (diamonds) based on the EMN~500 (open symbols) and EMN~450 (solid symbols) $NN$ interactions for different SVD ranks. Bottom panel: Relative error between rank-reduced HF and IMSRG(2) energies and their exact counterparts.}
    \label{fig:midmass}
\end{figure}

Figure~\ref{fig:SNM} shows the energy per particle of symmetric nuclear matter for densities up to $n = 0.2 \, \fm^{-3}$.
Again, we compare the full result against low-rank approximations with $\RSVD=1,\ldots,5$.
While the qualitative trend is reproduced at very low ranks as well there is a significant deviation from the exact result that becomes more pronounced with increasing density.
Once we include up to five singular values, the energy per particle is again quantitatively reproduced both at the HF level and at second-order MBPT.
While the energy per particle converges monotonically as a function of $\RSVD$ to the full results this is no longer true for applications in finite many-body systems as depicted in Fig~\ref{fig:midmass}.
The monotonicity should be seen as accidental in the nuclear matter case, since there is no variational principle imposing such convergence behavior.
Finally, we note that the absence of saturation is due to the lack of three-body forces in our study~\cite{Bogn05nuclmat,Hebe11fits}.

\section{Medium-mass nuclei}

To further demonstrate the power of the SVD and low-rank interactions, we consider as a final application the \ai{} description of medium-mass nuclei. The nuclear potentials are transformed from momentum space to a bound-state HO basis using 15 major shells. Subsequently, we perform an HF calculation followed by a non-perturbative in-medium similarity renormalization group (IMSRG)~\cite{Tsuk11IMSRG,Herg16PR,Hopp19medmass} calculation at the normal-ordered two-body level [IMSRG(2)] using the HF determinant as the reference state. In all calculations the HO frequency is set to $\hbar \Omega=28\, \MeV$.

In Fig.~\ref{fig:midmass} we compare the ground-state energy per particle at the HF and IMSRG(2) levels at various SVD ranks for selected closed-shell nuclei using the EMN~500 $NN$ interaction with cutoff $\Lambda = 500\, \MeV$ (open symbols) and $\Lambda=450 \, \MeV$ (solid symbols).
From the top panel we see that the ground-state energy per particle converges rapidly as a function of SVD rank and the full result is quantitatively reproduced when keeping five singular values in the decomposition in each partial-wave channel.
In most cases lower values of $\RSVD$ yield less binding both at the HF and IMSRG(2) level.
Although in all cases the full potential provides a bound HF solution, for $^{24}$O and $^{78}$Ni the HF determinant is unbound for the EMN~500 interaction at $\RSVD=1$.
This pathology is cured when increasing the decomposition rank.
We did not find a similar behavior when using the EMN~450 interaction.
Moreover, the quality of a low-rank approximation is stable as a function of mass number, i.e., the energy per particle is equally well reproduced at low ranks for the closed-shell oxygen and nickel isotopes.

A more careful analysis of the relative error (bottom panel) reveals that already at $\RSVD=3$ the decomposition error is only about three percent.
Further increasing to $\RSVD=5$ yields excellent reproduction of the exact results.
For the heaviest nucleus investigated so far the error on the IMSRG(2) binding energy for $\RSVD=5$ is $2.5 \, \MeV$ total.
For the systems studied here the relative error of IMSRG(2) ground-state energies seems to be smaller compared to the HF results at the same $\RSVD$.
However, the absolute error in both cases is comparable and the decrease in relative error reflects larger magnitude of the binding energy due to the additional correlations accounted for by the IMSRG solution.

While the decomposition was performed in momentum space, the various basis transformations may potentially spoil the low-rank properties of the tensor factorization. 
In particular, the basis transformations involving couplings of center-of-mass and intrinsic degrees of freedom may require the SVD to be reformulated.
However, since the SVD is performed in individual partial-wave channels, fundamental symmetries of the nuclear interaction, e.g., rotational invariance or parity conservation, are automatically preserved.
This is particularly important, since the conservation of spatial symmetries in a TF environment is highly non-trivial and more complex tensor formats will generally break symmetries of the many-body operators.

\section{Compression and computational scaling}

The key idea when following the TF paradigm is the lowering of computational resources at the price of inducing a small, controllable decomposition error.
The previous applications showed that for matrix dimension $N=100$ a decomposition rank of $\RSVD = 5$ is sufficient to obtain quantitative reproduction of nuclear matter and medium-mass nuclei based on chiral interactions.
While the number of matrix elements of a generic dense matrix scales quadratically with the matrix dimension, the SVD rank required for an accurate reproduction of observables is roughly of the order of only the square root of the basis size (matrix dimension) for all the cases considered in this work, i.e., $\RSVD \sim \sqrt{N}$ (note that we have focused on typical values of mesh sizes $N \sim 100$). Our applications showed that a compression of $\compression{100}{5}\approx 10$ can be obtained without significant loss in accuracy. 
We hope that similar low-rank properties will also exist in chiral three-nucleon ($3N$) interactions.
Taking typical $3N$ partial-wave dimensions of $N=5\cdot 10^4$~\cite{Hebe203NF}, compression ratios of $\compression{N}{\RSVD}>10^2$ could then be anticipated even for three-body partial-wave channels of moderate size, making it a promising tool to overcome current computational limitations.

While the three-body estimates certainly require validation using actual decompositions, the potential scaling reduction opens up new possibilities in designing low-rank nuclear-matter frameworks operating at higher truncation orders compared to their non-factorized counterparts, especially regarding higher-order MBPT $3N$ contributions.
These efforts nicely combine with recent developments in the automated derivation of such expressions that are, however, still beyond the scope of the presented factorizations~\cite{Arthuis2018adg1}.

\section{Conclusion}
\label{sec:conclusion}

In this Letter we systematically studied the impact of employing low-rank matrix factorizations of $NN$ interactions obtained from chiral effective field theory.
We found that singular values are rapidly suppressed across all partial-wave channels. Therefore, keeping only a very limited number of singular values provides an excellent approximation to the initial potential.
The quality of the low-rank potentials on two-body observables was confirmed in deuteron calculations and for various two-nucleon scattering phase shifts.
In the next step, the low-rank nuclear interactions were applied in perturbative many-body calculations of symmetric nuclear matter.
We observed that the energy per particle could be reproduced very well using rank-$5$ approximations for the potential.
Similarly, non-perturbative IMSRG calculations for closed-shell medium-mass nuclei confirmed the quality of low-rank potentials even after transforming to bound-state computational bases. All these benefits of singular value decompositions for nuclear interactions and many-body applications were recently confirmed in Ref.~\cite{Zhu2021}.

Based on these very promising results, we plan to extend the scope of tensor factorization techniques in various directions. 
A natural step is the application of SVD to nuclear three-body forces, with few-body benchmarks for light nuclei.
With this in hand, MBPT calculations with full inclusion of $3N$ interactions can be performed in nuclear matter.
Eventually, more aggressive factorization formats will be tested that will become important when targeting higher-mode tensors such as three-body matrix elements that depend on four Jacobi momenta.
All of these developments can potentially enable calculations of nuclei that have superior scaling with mass number.

\section*{Acknowledgements}

We thank Lars Zurek for useful discussions and comments on the manuscript, and Andreas Ekstr\"om for providing us with matrix elements for the $\Delta$-\nxlo{2}$_\text{GO}$ interaction.
This work was supported in part by the  Deutsche  Forschungsgemeinschaft (DFG,  German Research Foundation) -- Projektnummer 279384907 -- SFB 1245 and by the BMBF Contract No.~05P18RDFN1.

\bibliographystyle{apsrev4-1}
\bibliography{strongint}

\end{document}